\documentstyle[aps,preprint]{revtex}
\tightenlines
\begin{document}
\title{Approximations in Fusion and Breakup reactions induced by Radioactive
Beams}
\author{W.H.Z. C\'{a}rdenas$^1$, L.F.~Canto$^2$, R.~Donangelo$^2$,
M.S.~Hussein$^1$, J.~Lubian$^3$ and A. Romanelli$^4$}
\address{$^1$Instituto de F\'{\i}sica, Universidade de S\~{a}o Paulo,\\
C.P. 66318, 05389-970 S\~{a}o Paulo, Brazil\\
$^2$Instituto de F\'{\i}sica, Universidade Federal do Rio de Janeiro,\\
C.P. 68528, 21945-970 Rio de Janeiro, Brazil,\\
$^3$ Instituto de F\'\i sica, Universidade Federal Fluminense,\\
Av. Litor\^anea S-N, 24210-340 Niteroi, Rio de Janeiro, Brasil\\
and \\
CEADEN, Havana, Cuba\\
$^4$Instituto de F\'{\i}sica, Facultad de Ingenier\'\i a,\\
C.C. 30, Montevideo, Uruguay}
\date{\today}
\maketitle

\begin{abstract}
Some commonly used approximations for complete fusion and breakup
transmission coefficients in collisions of weakly bound projectiles at near
barrier energies are assessed. We show that they strongly depend on the
adopted classical trajectory and can be significantly improved with proper
treatment of the incident and emergent currents in the WKB approximation.
\end{abstract}


\section{Introduction}

In a fusion reaction the projectile and target nuclei form an excited
compound system, which decays by emission of particles or gamma rays. When
there are channels strongly coupled to the elastic channel, the reaction may
be described using any of the available coupled channels codes. The effects
of such couplings have been extensively discussed in the literature \cite%
{Da85} . In general terms, at energies below that of the Coulomb barrier,
these couplings tend to reduce the effective fusion barrier, which
substantially increases the fusion cross section. At high energies the
incident flux in the elastic channel is partially diverted into inelastic
and transfer channels. This tends to decrease the contribution of the
elastic channel to the fusion cross section, but this reduction is partially
compensated by the contribution from the other channels.

The recent availability of radioactive beams has made possible to study
reactions involving unstable nuclei. Such reactions are important in
processes of astrophysical interest, as well as in the search for superheavy
elements. The main new ingredient in reactions induced by unstable
projectiles is the strong influence of the breakup channel. In the case of
not too unstable projectiles, the effect of this channel in the fusion cross
section at low energies is, as in the case of stable beams, to enhance it.
At high energies, however, the situation is qualitatively different from the
case where only stable nuclei are involved. The contribution from the
breakup channel to the fusion reaction is strongly influenced by the low
probability that all fragments are captured. Thus, in this case, the fusion
cross section is partitioned into a complete and one or more incomplete
fusion contributions \cite{Ca98}.

The introduction of the breakup channel into a coupled channels calculation
is by no means trivial. The difficulty lies in the fact that this channel
lies in the continuum, and involves a, at least, three body system. This
problem has been addressed by several authors, using different approaches.
Several recent experiments involving fusion of neutron rich $^6$He and
proton rich $^{17}$F with heavy targets have been performed with the purpose
of exploring these theoretical proposals~\cite{Tr00,Ko98,Re98}. In Refs.~%
\cite{Ca98,Hu92,Hu93,Ta93,Ca95}, the coupled-channel problem is simplified
by the introduction of the polarization potentials arising from the coupling
with the breakup channel\cite{Ca92,AGN94}. In Refs.~\cite{Da94,Br95,Ha00},
the coupled-channel problem is solved directly within different
approximation, ranging from the schematic model of Dasso and Vituri \cite%
{Da94} to the huge calculation of Hagino {\it et al.}\cite{Ha00}, performed
through continuum discretization.

The polarization potential approach of Refs.~\cite{Ca98,Hu92,Hu93,Ta93,Ca95}
has the advantage of leading to simple expressions, which can easily be used
in data analysis\cite{Ta97}. However, it employs several approximations
which were not thoroughly tested. These approximations can be grouped in two
cathegories. In the first are those used in the derivation of the
polarization potentials. In the second are the semiclassical approximations
for fusion and breakup coefficients, used in calculations of the cross
sections. These coefficients are written in terms of barrier penetration
factors and survival probabilities, which are evaluated within the WKB
approximation. The aim of the present work is to ascertain the quality of
the approximations for the transmission coefficients. Approximations in the
derivation of the polarization potential will be the object of a latter
study. For our purposes, we consider a case where a complete quantum
mechanical calculation is feasible and compare exact and approximated cross
sections. We study the $^{11}$Li + $^{12}$C collision, using typical optical
and polarization potentials. For simplicity, our polarization potential has
no angular momentum or energy dependence and the range is given by the $^{11}
$Li breakup threshold energy. The strength is consistent with that found in
Ref.~\cite{Ca92} for the most relevant partial waves in near-barrier fusion.

The plan for this paper is as follows: in section II we briefly revise the
coupled channels formalism and the concept of a polarization potential. In
section III we discuss different approximation for the transmission
coefficients and investigate their consequences on the fusion and breakup
cross sections. Finally, in section IV, we present the conclusions of this
work.

\section{Coupled channel equations and polarization potentials}

In a standard coupled channels calculation, the system is described through
the distance between centers of projectile and target, {\bf r}, and a set of
intrinsic coordinates, $\xi$, that describe the internal degrees of freedom
of one of the nuclei, {\it e.g.} the target. These coordinates are
associated to an intrinsic Hamiltonian $h$ and its eigenfunction set,

\begin{equation}
h\phi _{\alpha }(\xi )=\epsilon _{\alpha }\phi _{\alpha }(\xi )\,,
\end{equation}
where

\begin{equation}
\int\phi_{\alpha}^{*}(\xi)\phi_{\beta}(\xi)d\xi= \delta_{\alpha,\beta}\,.
\end{equation}

The system Hamiltonian may then be written as

\begin{equation}
H=T+U^{opt}+h+v({\bf r},\xi )\,.
\end{equation}
Above, $T$ is the kinetic energy of the relative motion, $U^{opt}$ is the
optical potential and $v({\bf r},\xi )$ is the interaction coupling
intrinsic and collision degrees of freedom. The optical potential, which is
diagonal in channel space, accounts for the average interaction between
projectile and target.

Usually the solution of Schr\"{o}dinger's equation

\begin{equation}
H\Psi ({\bf r},\xi )=E\,\Psi ({\bf r},\xi )\,,  \label{schro1}
\end{equation}
where $E$\ is the collision energy in the center of mass frame, is expanded
as

\begin{equation}
\Psi ({\bf r},\xi )=\sum_{\alpha }\psi _{\alpha }({\bf r}) \phi
_{\alpha}(\xi )\,,  \label{chan-exp}
\end{equation}
where $\psi _{\alpha }({\bf r})$ describes the relative motion in channel $%
\alpha $. Substituting this expansion in Eq.~(\ref{schro1}) we obtain the
coupled channels equations (see e.g. Ref.~\cite{Sa83}),

\begin{equation}
(E_{\alpha }-H_{\alpha })\,\psi _{\alpha }({\bf r})= \sum_{\beta }{\cal V}%
_{\alpha \beta }({\bf r})\,\psi _{\beta }({\bf r})\,.  \label{CC}
\end{equation}
Above, $E_{\alpha }=E-\epsilon _{\alpha }$ and $H_{\alpha
}=T+U_{\alpha}^{opt}(r),$ where

\begin{equation}
U_{\alpha }^{opt}\equiv V_{\alpha }^{opt}-i\ W_{\alpha }^{opt}\,
\label{defopt}
\end{equation}
is the optical potential{\it \ }in channel{\it \ }$\alpha $. The imaginary
parts have the purpose of accounting for the flux lost to channels neglected
in the expansion of Eq.(\ref{chan-exp}). The channel coupling potentials, in
Eq.(\ref{CC}) are given by 
\begin{equation}
{\cal V}_{\alpha \beta }({\bf r})=\int d\xi \,\phi _{\alpha }^{\ast }
(\xi)\,v({\bf r},\xi )\,\phi _{\beta }(\xi )\,.  \label{V-alpha-beta}
\end{equation}

A consequence of the non-Hermitian nature of $H$ (see Eq.(\ref{defopt})) is
that the continuity equation breaks down. \ This can be checked following
the usual procedure to derive the continuity equation. \ For each $\alpha ,$
we evaluate $\psi _{\alpha }^{\ast }({\bf r})\times \left[ \text{Eq}.(\text{%
\ref{CC}})\right] \,-\, \left[ \text{Eq}.(\text{\ref{CC}})\right]
^{\ast}\times \psi _{\alpha }({\bf r})$ \thinspace and then sum the results.
Assuming that ${\cal V}_{\alpha \beta }$\ is hermitian, we obtain 
\[
\nabla \,\cdot \sum_{\alpha }\,{\bf j}_{\alpha }=\frac{2}{\hbar }
\sum_{\alpha }W_{\alpha }^{opt}({\bf r})\,|\psi _{\alpha }({\bf r})|^{2}
\neq 0 
\]
Integrating the above equation inside a large sphere with radius larger than
the interaction range and using the definition of the absorption cross
section, we obtain the useful relation\cite{Sa85}

\begin{equation}
\sigma _{a}=\frac{k}{E}\sum_{\alpha }\langle \psi _{\alpha }|\ W_{\alpha } \
| \psi _{\alpha }\rangle \ .  \label{sig_a-CC}
\end{equation}

\subsection{Polarization potentials}

In some coupled channel problems, it occurs that one is only interested in
the elastic wave function. One example is the study of complete fusion in
collisions involving nuclei far from stability, where the breakup threshold
is very low. An extreme example is $^{11}$Li, which has no bound excited
state. In such cases, the coupled channel problem involves only the elastic
and the breakup channels. Since the breakup channels contain at least three
fragments, their contribution to complete fusion is expected to be
negligible. Therefore, only the elastic wave function is required for the
calculation of the complete fusion and breakup cross sections.

In such cases, the {\it polarization potential} approach becomes very
convenient. It consists of replacing the coupled channel equations by a
single Schr\"odinger equation for the elastic state. This equation contains
a {\it polarization} term, $U^{pol}$, added to the optical potential and its
solution is identical to the elastic wave function obtained from the coupled
channel equations. According to Feshbach\cite{Fe62}, the polarization
potential is obtained through elimination of the coupled channel equations
for excited states and it is given by

\begin{equation}
U^{pol}=(\phi_0|PvQG^{(+)}_{QQ}QvP|\phi_0)\,.  \label{vpol}
\end{equation}

\noindent Above, $P=|\phi_0)(\phi_0|$ is the projector on the elastic
channel, $Q=1-P=\sum_{\alpha \ne 0}|\phi_{\alpha})(\phi_{\alpha}|$, and the
propagator $G^{(+)}_{QQ}$ is defined as

\begin{equation}
G^{(+)}_{QQ}=\frac{1}{E-QH_0Q + i\epsilon}\,.  \label{gr1}
\end{equation}

\noindent The wave function is then obtained by solving

\begin{equation}
(E-H_0-U^{pol})|\psi_0\rangle = 0\,,  \label{schopol}
\end{equation}

\noindent which, in the position representation is written

\begin{equation}
\left[ E-T-U^{opt}({\bf r})\right]\psi({\bf r})-\int U^{pol}({\bf r}, {\bf r}%
^{\prime})\:\psi({\bf r}^{\prime})d^3{\bf r}^{\prime}=0\,,
\end{equation}

\noindent where $U^{pol}({\bf r},{\bf r}^{\prime})$ is the nonlocal potential

\begin{equation}
U^{pol}({\bf r},{\bf r}^{\prime})= \sum_{\alpha}V_{0\alpha}({\bf r}
)\,G^{(+)}(E_{\alpha};{\bf r},{\bf r}^{\prime}) \,V_{\alpha 0} ({\bf r}%
^{\prime})\,.
\end{equation}

In principle, evaluating the polarization potential is nearly as hard as
solving the coupled channel equations. However, for practical purposes it is
replaced by trivially equivalent local potentials, which are calculated with
approximations\cite{Ca92,AGN94}.

\subsection{Fusion and breakup cross sections}

With the introduction of the polarization potential, any flux going away
from the elastic channel is treated as absorption. \ The sum in Eq.~(\ref%
{sig_a-CC}) is then reduced to a single term, the one with $\alpha =0.$ The
imaginary part of the potential is (henceforth we drop the superfluous index 
$\alpha $, since only the elastic channel appears), 
\begin{equation}
W=W^{opt}+W^{pol}\,,  \label{W-split}
\end{equation}
the absorption cross section can be split as 
\begin{equation}
\sigma _{a}=\sigma _{F}+\sigma _{bu}\,.  \label{sig-split}
\end{equation}
Above, 
\begin{equation}
\sigma _{F}=\frac{k}{E}\int d^{3}{\bf r}\ W^{opt}(r)\ |\psi ({\bf r})|^{2}
\label{sig-fusion}
\end{equation}
is identified with absorption through complete fusion and 
\begin{equation}
\sigma _{bu}=\frac{k}{E}\int d^{3}{\bf r}\ W^{pol}(r)\ |\psi ({\bf r})|^{2}
\label{sig-bup}
\end{equation}
corresponds to the loss of flux through the breakup channel. It includes the
breakup cross section and also a cross section for absorption in the breakup
channels, probably incomplete fusion. However, since for weakly bound nuclei
the range of $W^{pol}$ is much larger than that of $W^{opt},$ we neglect
this contribution and use the notaton $\sigma _{bu}$ in Eq.~(\ref{sig-bup}).

It is useful to consider the expansion in partial waves of the wavefunction,

\begin{equation}
\psi =\sum_{l,m}\frac{u_{l}(k,r)}{r}Y_{lm}(\theta ,\varphi )\,,  \label{onda}
\end{equation}
where $k=\sqrt{2\mu E/\hbar ^{2}}$ and the $u_{l}(k,r)$ are solutions of the
radial equation,

\begin{equation}
-\frac{\hbar ^{2}}{2\mu }\left[ \frac{d^{2}}{dr^{2}}-\frac{l(l+1)}{r^{2}} %
\right] u_{l}(k,r)+U^{opt}(r)\,u_{l}(k,r)=E\,u_{l}(k,r)\,,  \label{radial}
\end{equation}
normalized such that 
\begin{equation}
u_{l}(k,r\rightarrow \infty )=\frac{i}{2}\left[ H_{l}^{(-)}(kr)-S_{l}
\,H_{l}^{(+)}(kr)\right] \,.  \label{u-norm}
\end{equation}
Using the partial wave expansion in Eq.(\ref{sig-fusion}), the fusion cross
section may be rewritten as

\begin{equation}
\sigma _{F}=\frac{\pi }{k^{2}}\sum_{l}(2l+1)T_{l}^{F}\,,  \label{sigtl}
\end{equation}
where the transmission coefficient is given by

\begin{equation}
T_{l}^{F}=1-|S_{l}|^{2}=\frac{4k}{E}\int_{0}^{\infty }dr\ W^{opt}(r) \ |
u_{l}(k,r)|^{2}\ .  \label{tls}
\end{equation}
Proceeding similarly with Eq.(\ref{sig-bup}), we get 
\begin{equation}
\sigma _{bu}=\frac{\pi }{k^{2}}\sum_{l}(2l+1)T_{l}^{bu}\,,  \label{sigtl-bu}
\end{equation}
with

\begin{center}
\begin{equation}
T_{l}^{bu}=\frac{4k}{E}\int_{0}^{\infty }dr\ W^{pol}(r)\ |u_{l}(k,r)|^{2}\,.
\label{Tl-bu}
\end{equation}
\end{center}

\section{Approximations}

In what follows, we study different approximations to the coefficients $%
T_{l}^{F}$\ and $T_{l}^{bu}.$ In order to fix ideas, we consider a $^{11}$Li
beam incident on a $^{12}$C target, using an optical potential $%
U^{opt}=V^{opt}-i\ W^{opt}$ parameterized in the standard way:

\begin{equation}
V^{opt}(r)=V^{N}(r)+V^{C}(r)\,,
\end{equation}
with the nuclear part given by 
\begin{equation}
V^{N}(r)=\frac{V_{0}^{N}}{1+\exp \left[ \left( r-R_{r}\right) \,/a_{r} %
\right] }\,,  \label{vn}
\end{equation}
and the Coulomb one by

\begin{eqnarray}
V^{C}(r) &=&Z_{p}Z_{t}e^{2}/r;\quad \quad \quad \quad \quad \quad \quad
\quad \quad \quad {\rm for\;}r>R_{C} \\
&=&\left( Z_{p}Z_{t}e^{2}/2R_{C}\right) \ \left[ 3-\left( \frac{r}{R_{C}}
\right) \right] ^{2};\quad {\rm for\;}r\leq R_{C}\,.  \nonumber
\end{eqnarray}
Above, $Z_{p},A_{p}$ ($Z_{t},A_{t}$) are the atomic and mass numbers of the
projectile (target), $R_{C}$ is the radius of the nuclear charge
distribution, and $R_{r}$ is given by 
\begin{equation}
R_{r}=r_{r}^{0}\,\left( A_{p}^{1/3}+A_{t}^{1/3}\right) \,.  \label{R}
\end{equation}
The imaginary part is similarly parameterized as

\begin{equation}
W^{opt}(r)=\frac{W_{0}^{opt}}{1+\exp \left[ \left( r-R_{i}\right) \,/a_{i} %
\right] }\, ,  \label{wopt}
\end{equation}
\noindent with $R_{i}$ defined similarly to Eq.(\ref{R}). We take the
following parameter values:

\begin{equation}
V_0^{opt}=-60\ {\rm MeV};\quad r_r^0=1.25\ {\rm fm}; \quad a_r^0=0.60\ {\rm %
fm}\,;  \label{parV0}
\end{equation}

\begin{equation}
W_{0}^{opt}=60\ {\rm MeV};\quad r_{i}^{0}=1.00\ {\rm fm};\quad
a_{i}^{0}=0.60\ {\rm fm}\,.  \label{parW0}
\end{equation}

\bigskip Note that, since $W^{opt}$ corresponds exclusively to short range
fusion absorption, $r_{i}^{0}$ is appreciably smaller than $r_{r}^{0}.$

In order to review the standard approximations in the optical potential
calculations, we initially disconsider the breakup channels. In the absence
of breakup, the imaginary part of the nuclear potential has a short range,
and therefore fusion may be approximately described through an infinitely
absorbing imaginary potential with a well defined radius $R_{F}$. In this
case $T_{l}^{F}$ may be estimated by $T_{l}$, the transmission coefficient
through the effective potential

\begin{equation}
V_{l}(r)=V^{opt}(r)+\frac{\hbar ^{2}}{2\mu }\frac{l(l+1)}{r^{2}}\,.
\label{vl}
\end{equation}
If one approximates the region around the maximum of $V_{l}$ by a parabola,
then one obtains the Hill-Wheeler expression for $T_{l}^{F}$ \cite{HW53}

\begin{equation}
T_{l}^{F}\approx T_{l}\approx T_{l}^{HW}=\left\{ 1+\exp \left[ 2\pi \left( 
\frac{B_{l}-E}{\hbar \omega _{l}}\right) \right] \right\} ^{-1}\,,
\label{hw}
\end{equation}
where $R_{B}$ is the position of this maximum, $B_{l}$ its value, and $%
\omega _{l}$ the curvature of $V_{l}$ at $r=R_{B}$,

\begin{equation}
\hbar\omega_l=\left(-\ \frac{\hbar^2}{\mu}\left[\frac{d^2V_l(r)}{dr^2} %
\right]_{R_B}\right)^{1/2}.
\end{equation}
\bigskip

In Fig.~1, we show an example of a cross section calculated within the
Hill-Wheeler approximation (dashed line) compared with the exact quantum
mechanical calculation (full circles). One notices that the approximation is
excellent at energies above the Coulomb barrier, $E>V_{B}\equiv B_{l=0}$,
but worsens rapidly for $E<<V_{B}$.

The problem at low energies may be improved using the WKB approximation. The
transmission factor is then given by

\begin{equation}
T_{l}\approx \exp (-2\Phi )  \label{WKB}
\end{equation}
where

\begin{equation}
\Phi ={\rm Im}\left\{ \int_{r_{in}}^{r_{out}}k(r)dr\right\} \,.  \label{pore}
\end{equation}
Above,

\begin{equation}
k(r)=\frac{1}{\hbar }\ \sqrt{2\mu \ \left[ E-V_{l}(r)\right] }
\end{equation}
and $r_{in}$ e $r_{out}$ are the inner and outer classical turning points
for the potential $V_{l}$, determined through the condition $%
V_{l}(r_{in(out)})=E\,.$ However, this approximation is not good at energies 
$E\approx B_{l}$; for $E=B_{l}$ it yields $T_{l}=1$ instead of the correct
value $T_{l}=1/2$, and even worse, it does not predict reflections above the
barrier. Improvement is obtained by substituting the approximation of Eq.~(%
\ref{WKB}) by Kemble's expression \cite{Ke25} below the barrier while
keeping Hill-Wheeler's approximation above it,

\begin{equation}
T_{l}=(1+e^{2\Phi })^{-1}\;(E<B_{l});\quad \quad T_{l}=T_{l}^{HW}\; (E\geq
B_{l}).  \label{wkb1}
\end{equation}
We have employed this approximation in Ref.~\cite{Ca98}; it is equivalent to
employing the Hill-Wheeler formula for all energies, albeit with the
modification

\[
\frac{\pi }{\hbar \omega _{l}}(B_{l}-E)\longrightarrow \Phi ;\quad {\rm for}
\;E>B_{l}\,. 
\]
The cross section obtained within this approximation is depicted in Fig.~1
(solid line). We see that it reproduces the full quantum calculations for
all collision energies.

Let us now consider the inclusion of the breakup channels. As we have seen,
this may be done through the introduction of an appropriate polarization
potential. \ Such potentials were studied in Refs.~\cite{Hu93,Ta93}, for
pure nuclear coupling, and in \cite{Ca95,AGN94} for the electromagnetic
coupling. \ In \cite{Hu93,Ta93} only the imaginary part of the polarization
potential was calculated. Since the real part of the polarization potential
reduces the height of the potential barrier, this effect was simulated by a
shift in the collision energy in the calculation of $T_{l}$. Namely, 
\[
T_{l}(E)\rightarrow T_{l}(E+\Delta E);\quad \Delta E=-\,V^{pol}(R_{B})\,. 
\]
\ As we will see, the real part of the polarization potential plays a very
important role at energies below the Coulomb barrier. In the case of $^{11}$%
Li + $^{12}$C, the breakup process is dominated by the nuclear coupling.
Therefore we write

\begin{equation}
W^{pol}(r)=\frac{W_{0}^{pol}(l,E_{CM})}{1+\exp \left[ \left(
r-R_{pol}\right) /\alpha \right] }\,,  \label{wpol}
\end{equation}
where $R_{pol}$ may be approximated by the optical potential radius, and the
diffuseness $\alpha $ is given in terms of the breakup threshold energy $%
B_{bu}$ as

\begin{equation}
\alpha =\left( \frac{2\mu _{bu}B_{bu}}{\hbar ^{2}}\right) ^{1/2}\,.
\label{alpha}
\end{equation}
Above, $\mu _{bu}$ is the reduced mass of the fragments produced in the
breakup process. In the case of $^{11}$Li, $B_{bu}=0.2$~MeV and thus $\alpha
=6.6$~fm.

The strength of the polarization potential varies with $l$ and $E_{CM}$,
and, for the partial waves relevant to the fusion process, is of the order
of 1~MeV in the region around $r\approx R_{pol}$. Since in this work we are
not concerned with its derivation, but with the approximations employed in
the determination of the cross section, we shall adopt the constant value

\begin{equation}
W_{0}^{pol}(l,E)\equiv W_{0}^{pol}=2.0\,{\rm MeV}\,.
\end{equation}

Since the real part of the polarization plays a very important role at
energies below the Coulomb barrier, we shall include it here. In the
calculations of Andr\'es {\it et al.} \cite{AGN94} the real and imaginary
parts of the polarization potential have qualitatively the same strengths.
For simplicity we then take them to be equal, {\it i.e.}

\begin{equation}
V_{0}^{pol}(l,E)\equiv V_{0}^{pol}=-2.0\,{\rm MeV}\,.
\end{equation}

The effect of the real and imaginary parts of the polarization potential are
shown in Fig.~2. As it could be expected, the real part leads to a
substantial increase in the fusion cross section, most evident at energies
below the Coulomb barrier. On the other hand, the imaginary part reduces the
cross section both above and below the barrier. When both the real and
imaginary parts are included, there is a competition between the effects of
the real and imaginary parts. With the polarization strength values
considered above, suppression dominates above the barrier and enhancement
below it. This situation was also encountered in the coupled channels
calculations of Breitschaft {\it et al.}\cite{Br95} and Hagino {\it et al.}~ %
\cite{Ha00}.

The presence of a long-ranged absorption requires the introduction of
modifications in the approximations to $T_{l}^{F}$. Now the flux that
reaches the strong absorption region is attenuated not only because of the
reflection at the barrier, but also because of its absorption into the
breakup channel. In Ref.~\cite{Hu92} it was proposed the approximation

\begin{equation}
T_{l}^{F}\approx T_{l}(E+\Delta E)\cdot P_{l}^{surv}\,,  \label{psurv}
\end{equation}
where $T_{l}(E+\Delta E)$ is the WKB transmission factor (Eq.~(\ref{WKB}))
evaluated at the energy $E+\Delta E$\ and $P_{l}^{surv}$ is the breakup
survival probability. Within the WKB approximation we may take

\begin{equation}
P_{l}^{surv}=\exp \left[ -\frac{2}{\hbar }\int \frac{W^{pol}(r)}{v_{l}(r)}dr %
\right] \,,  \label{plsob}
\end{equation}
where $v_{l}(r)$ is the local radial velocity along a classical trajectory
with angular momentum $\hbar l$. \ A more formal justification for Eq~(\ref%
{psurv}), based on a WKB calculation with three turning points was presented
in Ref.~\cite{Ta93}. \ This approximation is consistent with the results of
Fig.~2. \ The enhancement due to $V^{pol}$ is incuded in $T_{l}$ while the
suppression arising from $W^{pol}$ is contained in $P_{l}^{surv}.$

In order to estimate $P_{l}^{surv}$ one needs to define the classical
trajectories to be employed in the calculation. In Ref.~\cite{Hu92} we
considered pure Rutherford trajectories, neglecting the nuclear potential
diffractive effects. These trajectories present a single turning point. The
corresponding fusion cross section is shown in Fig.~3 as a thin line with
solid circles. This figure also depicts the full quantum mechanical results
(thick solid line). We see that although the approximation obtained with the
Rutherford trajectory is reasonable at high energies, it breaks down at
energies close and below the Coulomb barrier ($V_{B}=2.67$~MeV). The
inclusion of the nuclear potential in the trajectory calculations improves
considerably the results (thin line with stars). In this case we may have,
depending on the partial wave and collision energy, one or three turning
points. The treatment with three turning points is not accurate in the
region around the Coulomb barrier, and that is the reason why there are
large deviations in the approximated fusion cross section. Later we will
show how one may improve this approximation, but let us first briefly
consider the breakup cross section.

In Ref.~\cite{Hu92} the breakup was calculated by considering that

\begin{equation}
T_{l}^{bu}=1-P_{l}^{surv}\,.  \label{pbu-0}
\end{equation}
This approximation is based on the notion that $T_{l}^{bu}$ corresponds to
the probability of non-survival to the breakup process. The results depend
strongly on the classical trajectory considered. In Fig.~4 we compare the
exact quantum mechanical breakup cross section to the ones obtained using
Eq.~(\ref{pbu-0}) with different trajectories. The results are far from
satisfactory. In particular, when the nuclear potential is included in the
trajectory calculations the low energy breakup cross section has a
completely wrong behavior. The reason for this discrepancy has been
discussed by Takigawa {\it et al.} \cite{Ta93} and will be considered in
further detail later in this section.

Let us now develop an improved WKB approximations for $T_{l}^{F}$ and $%
T_{l}^{bu}$. In order to explain them, it will be useful to rewrite the $%
T_{l}^{F}$ coefficients in a different way. In the WKB approximation, the
radial wave funcions with incoming $(-)$ and outgoing $(+)$ boundary
conditions are given by

\begin{equation}
u_{l}^{(\pm )}(r)=\frac{A}{\sqrt{k(r)}}\exp \left[ \pm \,i\int dr\,\,k(r) %
\right] \,,  \label{wkb2}
\end{equation}
where

\begin{equation}
k(r)=\frac{1}{\hbar }\sqrt{2\mu \left[ E-U^{opt}(r)-\frac{\hbar ^{2}}{2\mu
r^{2}}l(l+1)-U^{pol}(r)\right] }\,.  \label{k}
\end{equation}

The value of $T_{l}^{F}$ is given by the ratio between the probability
density current that reaches the strong absorption region, $j^{(-)}(r=R_{F})$%
, to the incident one $j^{(-)}(r=\infty )$, where the radial currents are 
\begin{equation}
j^{(\pm )}(r)=\frac{\hbar }{2\mu i}\,\left[ \left( u_{l}^{(\pm )}(r) \right)
^{\ast }\,\left( \frac{du_{l}^{(\pm )}(r)}{dr}\right) -\,
u_{l}^{(\pm)}(r)\,\left( \frac{du_{l}^{(\pm )}(r)}{dr}\right) ^{\ast }\right]
\,.  \label{j+-}
\end{equation}
From Eqs.(\ref{wkb2}) to (\ref{j+-}), we obtain

\begin{equation}
T_{l}^{F}=\frac{j^{(-)}(r=R_{F})}{j^{(-)}(r=\infty )}\approx \exp \left[ -2 
\bar{\Phi}\right] \,,  \label{atenuation}
\end{equation}
where

\begin{equation}
\bar{\Phi}={\rm Im}\left\{ \int_{R_{F}}^{\infty }dr\,k(r)\right\} \,.
\label{Phi-at}
\end{equation}

If one does not include the polarization potential, the integrand in the
equation above is real on the whole classically allowed region (note that $%
W^{opt}(r>R_{F})=0$). In this way, only the classically forbidden region
contributes to attenuate the current that reaches the fusion region ($%
r<R_{F} $), {\it i.e.}

\begin{equation}
\bar{\Phi}\rightarrow \Phi =\int_{r_{in}}^{r_{out}}dr\,k(r)\,,
\end{equation}
where $r_{in}$ and $r_{out}$\ are the inner and outer turning points. In
this case $T_{l}^{F}$ reduces to the expression given in Eq.~(\ref{WKB}).

However, if there is long-ranged absorption as a result of the coupling to
the breakup channels, the integrand in Eq.~({\ref{Phi-at}}) becomes complex
in all the integration region. The contributions to the integral that
defines $\bar{\Phi}$ from the classically allowed and forbidden regions may
be calculated separately. In this case, $T_{l}^{F}$ is written as the
product of factors resulting from each of them. Disregarding the imaginary
part of $U^{pol}(r)$\ in the classically forbidden region, the corresponding
factor reduces to the WKB tunneling probability $T_{l}$. On the other hand,
in the classically allowed regions $k(r)$ can be calculated in an
approximate way. Assuming that the imaginary part of $U^{pol}(r)$ is small
in comparison to the remaining terms in the square root appearing in Eq.~(%
\ref{k}), we may take a series expansion to the lowest order,

\begin{equation}
k(r)\simeq k_{0}(r)+i\ \frac{W^{pol}(r)}{\hbar \,v(r)}\,,
\end{equation}
where

\begin{equation}
k_{0}(r)=\frac{1}{\hbar }\sqrt{2\mu \left[ E-U^{opt}(r)-\frac{\hbar ^{2}} {%
2\mu r^{2}}l(l+1)-V^{pol}(r)\right] }\,
\end{equation}
and $v(r)$ is the local velocity,

\begin{equation}
v(r)=\frac{\hbar k_{0}(r)}{\mu }\,.
\end{equation}
Since $k_{0}(r)$ does not attenuate the incident probability current, we
obtain the same factor $P_{l}^{surv}$ as before.

In our procedure we do not explicitly distinguish between classically
allowed and forbidden regions, and calculate $\bar{\Phi}$\ directly from
Eq.~(\ref{Phi-at}), without any of the additional approximations mentioned
in the previous paragraph. In Fig.~5 we show the fusion cross section
obtained within this approximation, compared with the exact results and with
the old approximation. We see that the present approximation yields
excellent results in all energy regions, including the one around the
Coulomb barrier where the old approximation totally failed.

As noted by Takigawa {\it et al.} \cite{Ta93}, the relationship between $%
T_{l}^{bu}$ and $P_{l}^{surv}$ that appears in Eq.~(\ref{pbu-0}) is not
actually correct. The reason for this is that when we calculate the survival
probability we consider only the incident branch of the trajectory, along
which the system approaches the strong absorption region. However, the
breakup process may take place both on the entrance or exit branches. Let us
first consider the calculation of Ref.~\cite{Hu92}, which determines $%
P_{l}^{surv}$ along a Rutherford trajectory. The survival probability
associated with $T_{l}^{bu}$ is the one calculated along the whole
trajectory, {\it i.e.} along both branches A and B in Fig.~6a, and not just
along branch A, as it was done in the calculation of $P_{l}^{surv}$. Since
the contribution from both branches to the integral that defines $\bar{\Phi}$
(Eq.~(\ref{Phi-at})) are equal, the breakup probability amplitude may be
written as

\begin{equation}
T_l^{bu}=1-\left(P_l^{surv}\right)^2\,.  \label{Tbu-0}
\end{equation}

If we now take into account the effect of the nuclear potential on the
classical trajectory, the situation changes very much. For low partial
waves, where $E>B_{l}$, the infinite absorption condition in the strong
absorption region allows for only an ingoing branch. On the other side, for
partial waves for which $E<B_{l}$ we may have two classical turning points,
as illustrated in Fig.~6b. In that case all segments A, B, C do contribute
to the breakup cross section. In this case the amplitude $T_{l}^{bu}$ is
given by

\begin{equation}
T_{l}^{bu}=\frac{\left[ j_{l}^{(-)}(\infty )-j_{l}^{(-)}(r_{out})\right]
\,+\,\left[ j_{l}^{(-)}(r_{in})-j_{l}^{(-)}(R_{F})\right] \,+\,\left[
j_{l}^{(+)}(r_{out})-j_{l}^{(+)}(\infty )\right] }{j_{l}^{(-)}(\infty )}\,.
\label{tlbu1}
\end{equation}
The first term in the numerator corresponds to the contribution to the
breakup channel along incoming branch A in Fig.~6b. The second term
corresponds to the other incoming segment, C, while the third one is the
contribution associated to the exit branch B. The currents in this equation
are given by

\begin{eqnarray}
j_{l}^{(-)}(r_{out}) &=&e^{-2\Phi _{1}}\,\,j_{l}^{(-)}(\infty )  \nonumber \\
j_{l}^{(-)}(r_{in}) &=&T_{l}\,\,j_{l}^{(-)}(r_{out})  \nonumber \\
j_{l}^{(-)}(R_{F}) &=&e^{-2\Phi _{2}}\,\,j_{l}^{(-)}(r_{in})  \nonumber \\
j_{l}^{(+)}(r_{out}) &=&(1-T_{l})\,\,j_{l}^{(-)}(r_{out})  \nonumber \\
j_{l}^{(+)}(\infty ) &=&e^{-2\Phi _{1}}\,\,j_{l}^{(+)}(r_{out})\,,
\end{eqnarray}%
where $\Phi _{1}$ e $\Phi _{2}$ are given by

\begin{equation}
\Phi _{1}={\rm Im}\left\{ \int_{r_{out}}^{\infty }\,dr\,k(r)\right\}
\,;\quad \Phi _{2}={\rm Im}\left\{ \int_{r_{in}}^{r_{out}}\,dr\,k(r)\right\}.
\label{Phi1-Phi2}
\end{equation}
Substituting the density currents in Eq.~(\ref{tlbu1}), we obtain

\begin{equation}
T_l^{bu}=\left[ 1-e^{-2\Phi_1}\right] \,+\,e^{-2\Phi_1} \left[ T_{l}\,
\left(1-e^{-2\Phi_2}\right) \,+\,\left( 1-T_{l}\right) \,
\left(1-e^{-2\Phi_{1}}\right) \right] \,.  \label{Tbu2}
\end{equation}

The breakup cross section calculated using Eqs.~(\ref{Tbu-0}) (dashed line)
and (\ref{Tbu2}) (full line) are shown in Fig.~7, where they are compared to
exact results (solid circles). We notice that the two approximations lead to
similar results, and both are reasonably close to the exact values.
Comparing the two curves we reach two important conclusions. One is that the
inaccuracy in the results in Fig.~4 is due to the omission of the exit
branch in the trajectories. The other is that in the present case nuclear
effects on the trajectory are not very relevant. This is because the most
important contributions to the breakup cross section arise from the high-$l$
partial waves. While for the energy range considered the fusion cross
section converges for $l=10$, the breakup one requires the inclusion of
partial waves as high as $l\approx 80$. In this way, for most partial waves
relevant for the breakup calculation the external turning point is placed
outside the nuclear potential range. The situation changes somewhat when a
more realistic potential is considered. In that case, its intensity
decreases at high $l$ values, and the breakup cross section becomes more
sensitive to low partial waves.

\section{Conclusions}

We have investigated the validity of commonly used approximations for
complete fusion and breakup transmission coefficients in collisions of
weakly bound projectiles at near barrier energies. They were tested in the
concrete example of the of a $^{11}$Li + $^{12}$C collision. For the
calculations, we adopted a typical strong absorption optical potential and,
for simplicity, a schematic polarization potential, consistent with
theoretical predictions available in the litterature\cite{Ca92,AGN94}.

We have shown that the factorization of the complete fusion transmission
coefficient as a tunneling factor times a survival probability \cite%
{Ca98,Hu92,Hu93,Ta93,Ca95} may be a reasonable approximation, depending on
the classical trajectory used in the evaluation of the latter. However, it
is always inaccurate in the neighbourhood of the Coulomb barrier. An
improved WKB approximation for $T_{l}^{F}$ was shown to lead to very
accurate values of the complete fusion cross section in the whole energy
range of our study, both above and below the Coulomb barrier.

We have also shown that the breakup transmission coefficient can be obtained
in terms of survival probabilities provided that the emergent branch of the
classical trajectory is included in the calculation. However, the accuracy
of this approximation is worse than that for complete fusion.

\vspace{0.5in}

This work was supported in part by CNPq and the MCT/FINEP/CNPq(PRONEX) under
contract no. 41.96.0886.00. L.F.C. and R.D. acknowledge partial support from
the Funda\c{c}\~{a}o Universit\'{a}ria Jos\'{e} Bonif\'{a}cio, and M.S.H.
and W.H.Z.C. acknowledge support from the FAPESP.

\vspace{0.5cm}

\bigskip

{\bf Figure Captions}

\begin{itemize}
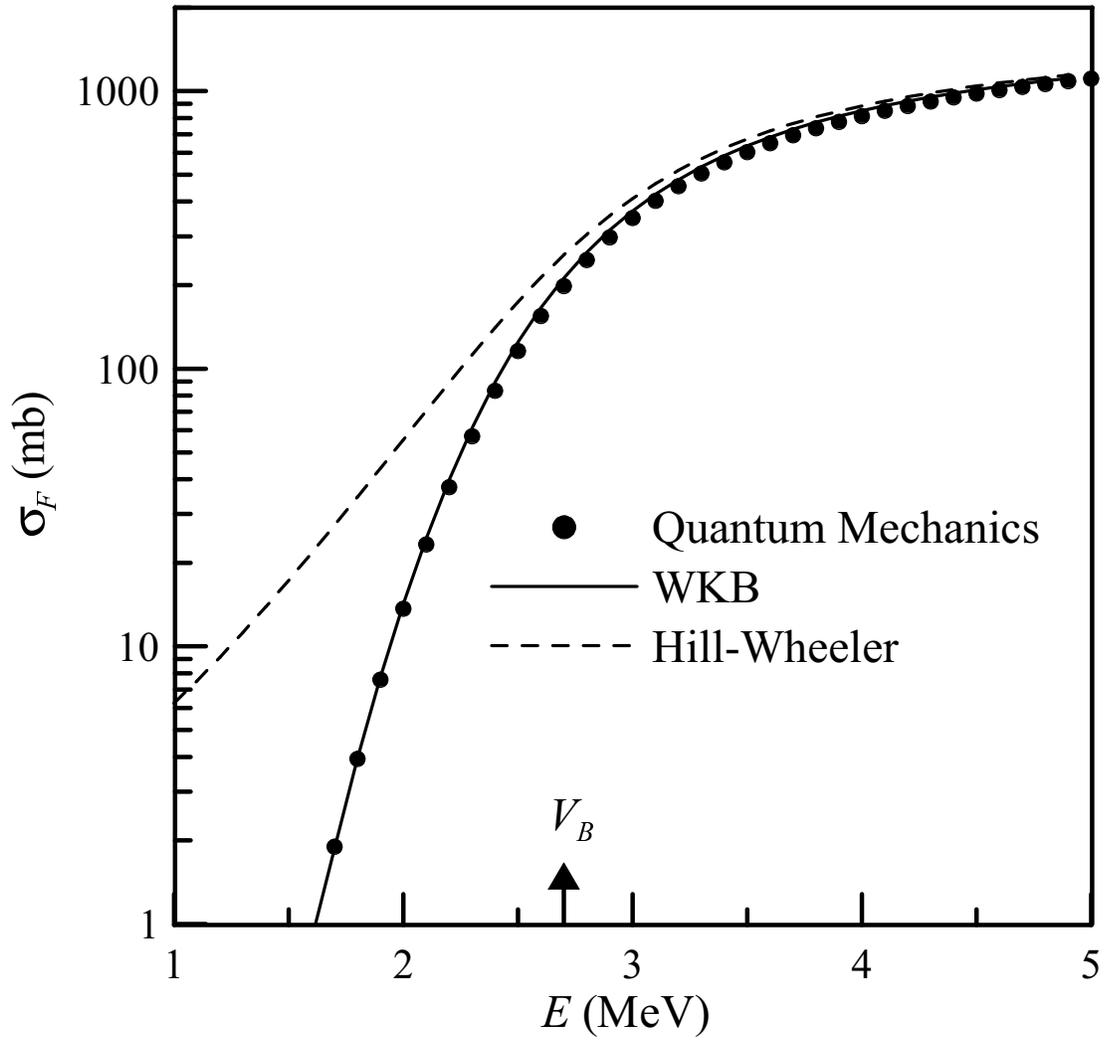
\item Figure 1: Hill-Wheeler and WKB approximations to the fusion cross
section. The vertical arrow indicates the position of the Coulomb barrier.
See text for further details.

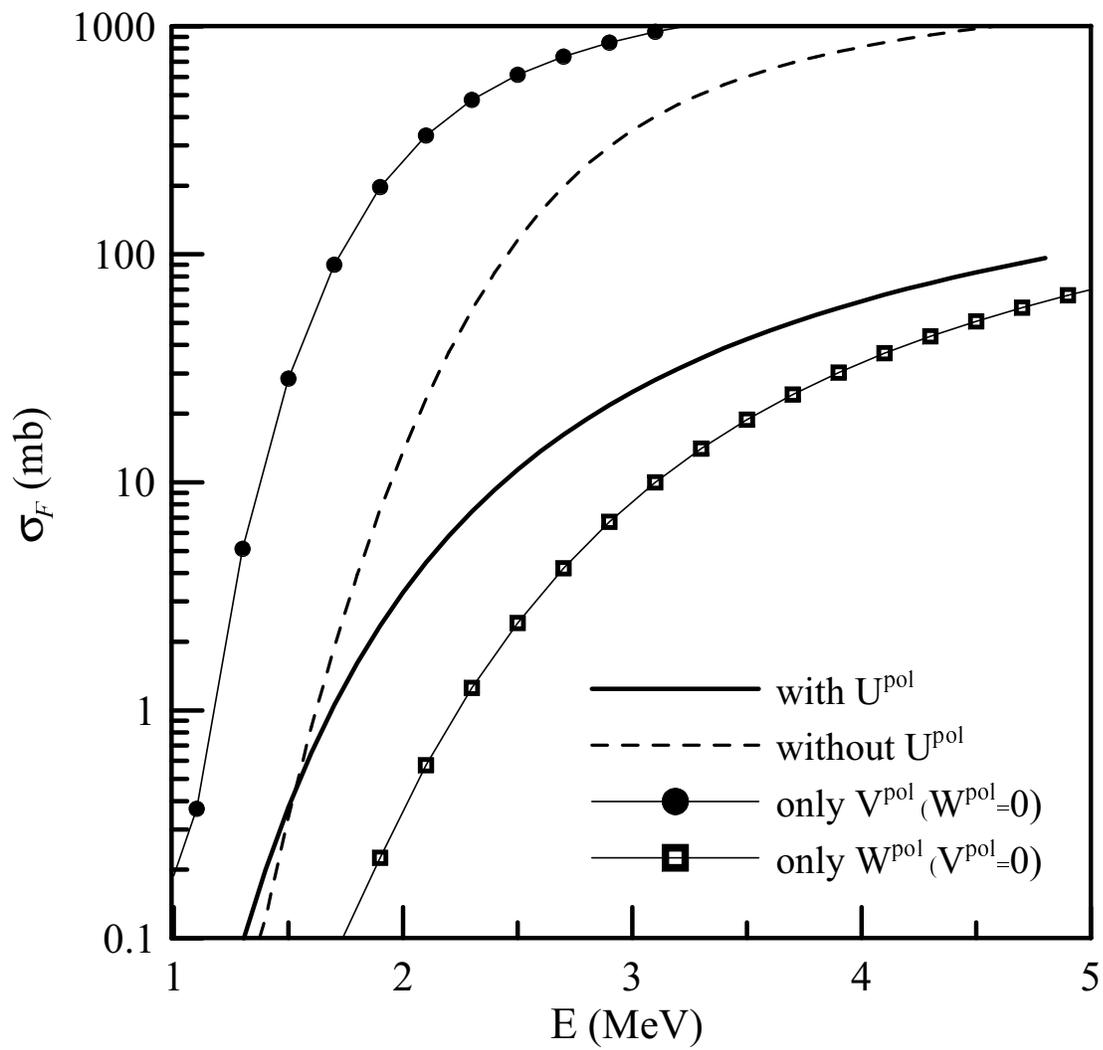
\item Figure 2: Fusion cross section with different contributions of the
polarization potential. See text for details.

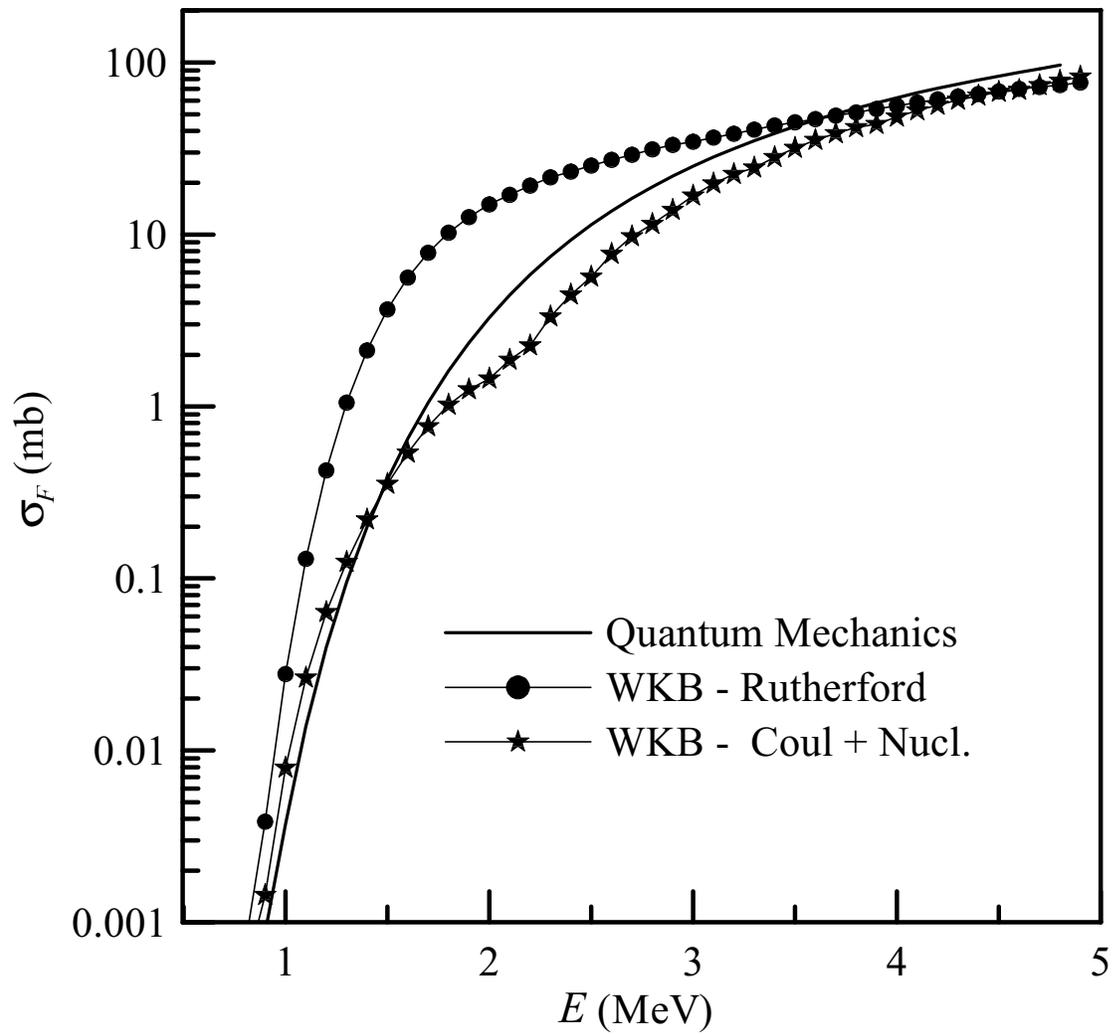
\item Figure\ 3: Fusion cross sections obtained with different
approximations employed in previous publications. The solid line indicates
exact quantum mechanical calculations and the remaining ones are obtained
with survival probability approximation (Eq.~(\ref{psurv})). The solid
circles were obtained with Rutherford trajectories while the starts were
obtained with classical trajectories taking into account both the Coulomb
and the nuclear potentials.

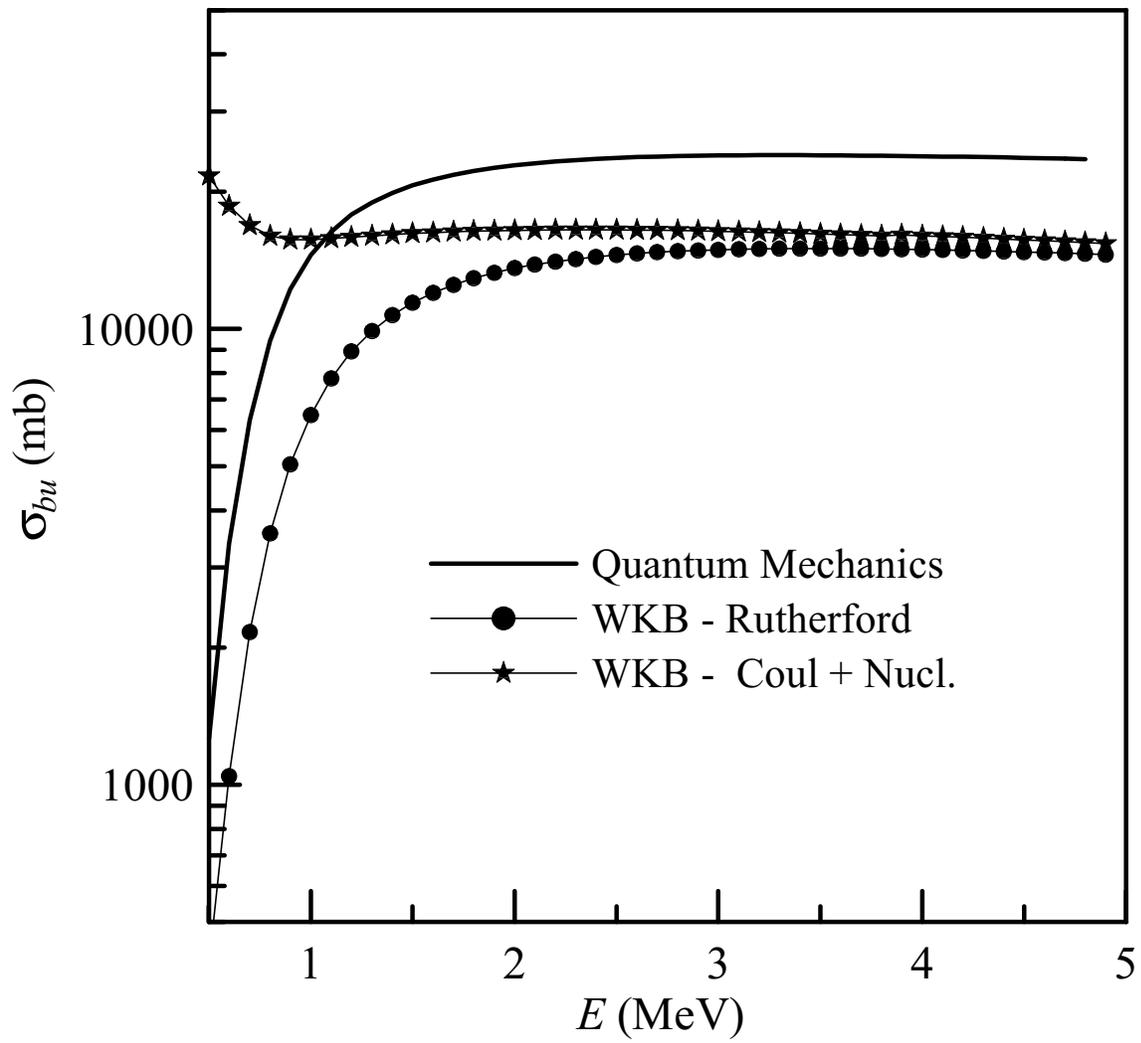
\item Figure~4: Exact breakup cross section (solid line) and cross sections
approximated by Eq.~(\ref{pbu-0}). The solid circles were obtained with pure
Rutherford trajectories while the stars takes into account nuclear potential
effects on the trajectory.

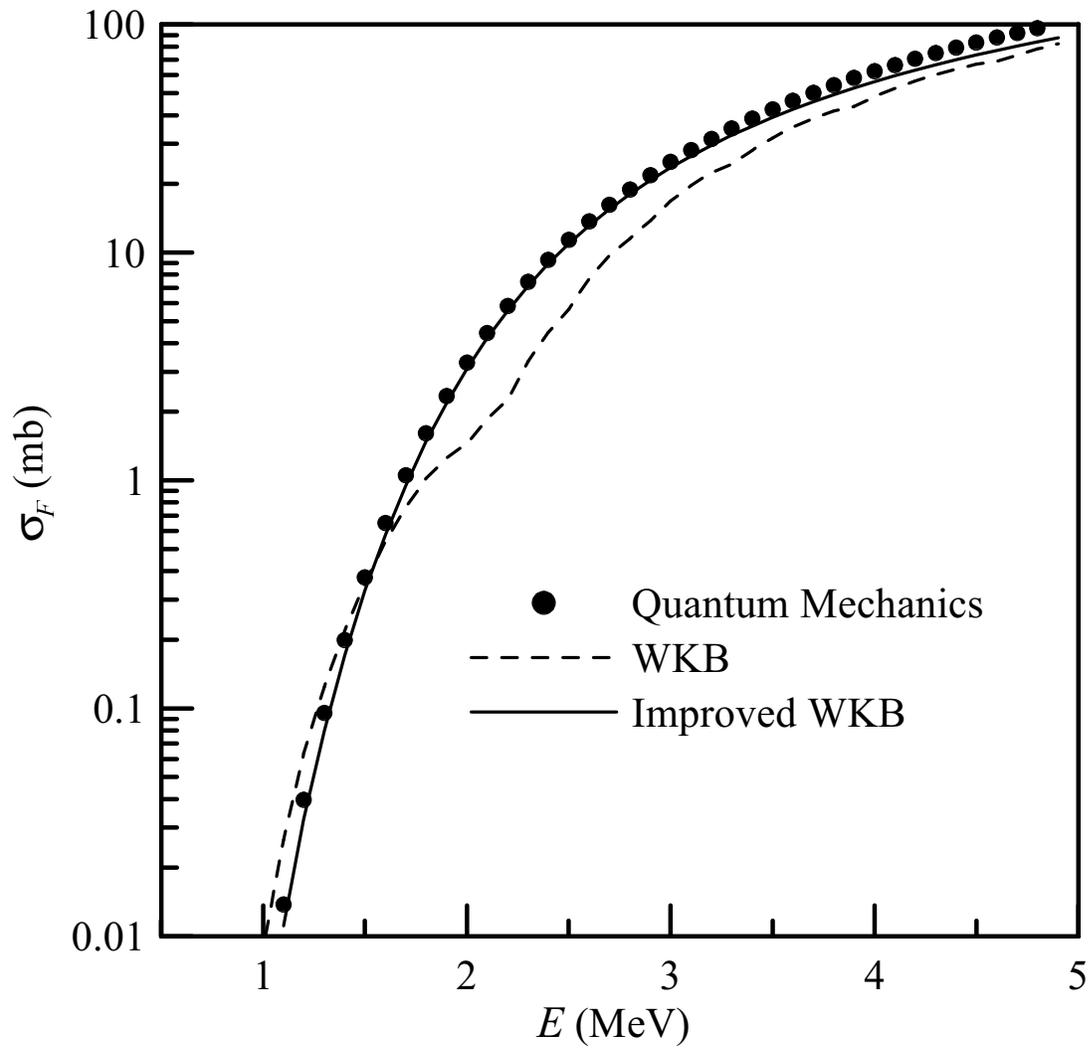
\item Figure~5: Exact complete fusion cross section (solid circles) compared
to the old approximation, depicted also in Fig.~3 (stars), and with the
improved WKB approximation (solid line). See text for more details.

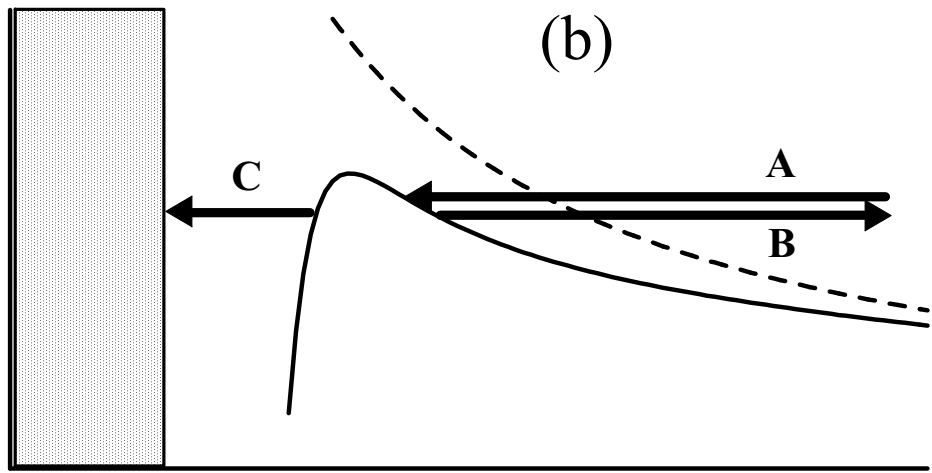
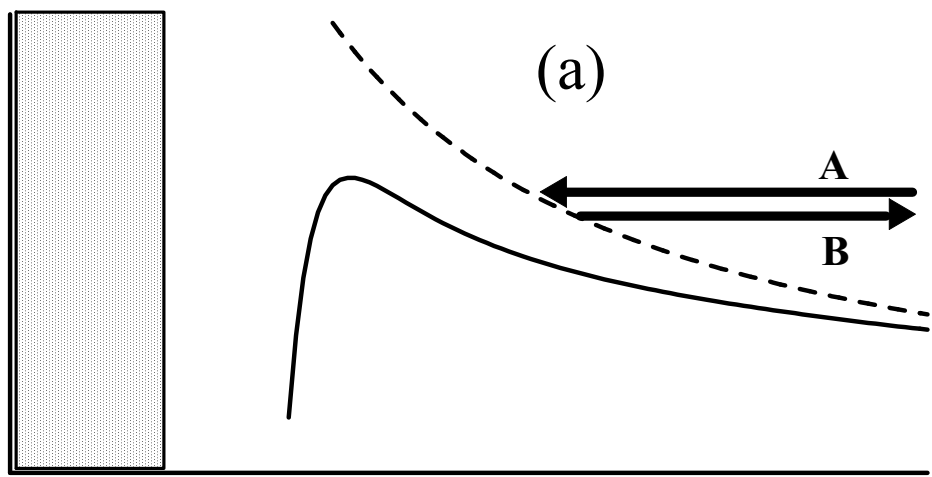
\item Figure 6: Branches (A, incoming; B, outgoing) of the collision
trajectory that contribute to the breakup process, (a) pure Rutherford, and
(b) including nuclear potential effects. In this later case, the incoming
branch has an additional segment (C).

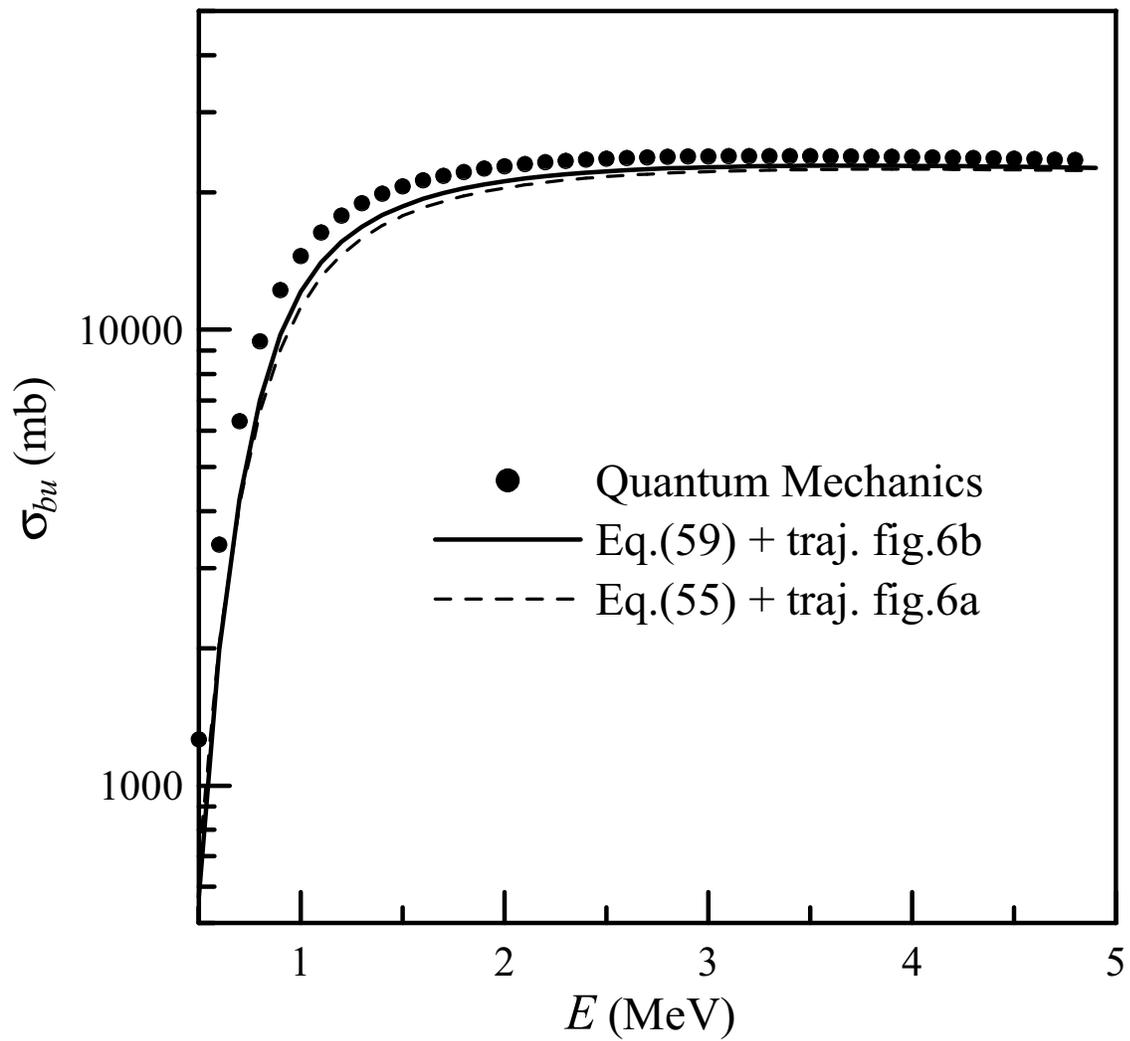
\item Figure~7: Exact calculations of the breakup cross section (solid
circles) compared to WKB calculations taking into account all branches of
the classical trajectory (A, B, C in Fig.~6b and Eq.~(\ref{Tbu2})) (solid
line) and taking only branches A and B in the Rutherford trajectory (Fig.~6a
and Eq.~(\ref{Tbu-0})) (dashed line).
\end{itemize}

\end{document}